\documentclass{PoS}

\usepackage{amsmath}

\newcommand{\id}{\mathrm{d}}
\newcommand{\nn}{\nonumber}

\newcommand{\define}{\equiv}
\newcommand{\tekst}{\textrm}

\newcommand{\Li}{\tekst{Li}}

\newcommand{\Litt}{\Li_{2,2}}
\newcommand{\floor}[1]{\left \lfloor #1 \right \rfloor}

\DeclareFontFamily{U}{wncy}{}
\DeclareFontShape{U}{wncy}{m}{n}{<->wncyr10}{}
\DeclareSymbolFont{mcy}{U}{wncy}{m}{n}
\DeclareMathSymbol{\sha}{\mathord}{mcy}{"58}

\title{On the Reduction and Evaluation of Generalized Polylogarithms}

\ShortTitle{Reduction and Evaluation of GPLs}

\author{\speaker{Hjalte Frellesvig}\\
        Institute of Nuclear and Particle Physics, NCSR ``Demokritos'',\\
Patriarchou Gregoriou E. \& Neapoleos 27, Agia Paraskevi, 15310, Greece\\
        E-mail: \email{frellesvig@inp.demokritos.gr}}

\author{Damiano Tommasini\\
Institute of Nuclear and Particle Physics, NCSR ``Demokritos'',\\
Patriarchou Gregoriou E. \& Neapoleos 27, Agia Paraskevi, 15310, Greece\\
        E-mail: \email{tommasini@inp.demokritos.gr}}

\author{Christopher Wever\\
Institute of Nuclear and Particle Physics, NCSR ``Demokritos'',\\
Patriarchou Gregoriou E. \& Neapoleos 27, Agia Paraskevi, 15310, Greece \& \\
Institute for Theoretical Particle Physics (TTP), Karlsruhe Institute of Technology,\\
Engesserstra{\ss}e 7, D-76128 Karlsruhe, Germany \& \\
Institute for Nuclear Physics (IKP), Karlsruhe Institute of Technology,\\
Hermann-von-Helmholtz-Platz 1, D-76344 Eggenstein-Leopoldshafen, Germany\\
        E-mail: \email{christopher.wever@kit.edu}}

\abstract{The talk and the summary below was based on the paper ``On the reduction of Generalized Polylogarithms to $\Li_n$ and $\Litt$ and on the reduction thereof'' \cite{Frellesvig:2016ske} by the three authors, published in March 2016 in JHEP.}

\FullConference{Loops and Legs in Quantum Field Theory\\
		24-29 April 2016\\
		Leipzig, Germany}

\begin{document}

\section{Introduction}
\label{sec:intro}

The Generalized Polylogarithmic function (GPL) \cite{Goncharov:2011ar, Goncharov:2010jf} is defined recursively as
\begin{align}
G(a_1,\ldots,a_n;x) &= \int_0^x \! \frac{\id z}{z-a_1} G(a_2,\ldots,a_n;z)\,,
\label{eq:gdef}
\end{align}
with
\begin{align}
G(\underbrace{0,\dots,0}_{n};x) \,=\, \frac{\log^n(x)}{n!} \;\;\;\;\;\;\;\;\;\; \tekst{and} \;\;\;\;\;\;\;\;\;\; G(;x) \,=\, 1\,,
\label{eq:gdefzeroes}
\end{align}
with the integration path being a straight line from $0$ to $x$.
This function generalizes a large number of other functions, such as the logarithm $\log(x)$, the classical polylogarithm $\Li_n(x)$, and the harmonic polylogarithm $H_{\bar{m}}(x)$ \cite{Remiddi:1999ew}.

At the one-loop order, all Feynman integrals may be expressed in terms of GPLs, and at two and more loops this remains true for a large class of the Feynman integrals, see e.g. the reviews \cite{Duhr:2014woa,Henn:2014qga}.

The utility of GPLs may also be appreciated from a purely mathematical point of view, since an integral of elemental functions, such as
\begin{align}
I &= \int_0^x \frac{\log(t) \log (1-t) \log(1 - t/a)}{t-b} \id t
\end{align}
has no result expressible in terms of more standard functions, but if results in terms of GPLs are allowed for, the result may be expressed as
\begin{align}
I \, &= \, G(b,0,1,a;x) + G(b,0,a,1;x) + G(b,1,0,a;x) \nn \\
&\; + G(b,1,a,0;x) + G(b,a,0,1;x) + G(b,a,1,0;x).
\end{align}
These considerations should motivate the further investigation of GPLs.

Many relations among GPLs are known. Primary are the rescaling relation
\begin{align}
G(a_1,\ldots,a_n;x) &= G(z a_1,\ldots, za_n; zx) \;\;\;\;\;\; \tekst{with} \;\;\;\;\;\; a_n,z \neq 0
\label{eq:rescaling}
\end{align}
and the shuffle relation
\begin{align}
G(a_1,\ldots,a_m;x) G(b_1, \ldots, b_n; x) &= \!\!\!\!\! \sum_{c \, \in \, a \sha b} \!\!\!\!\! G(c_1, \ldots, c_{m+n}; x)\,,
\label{eq:shuffle}
\end{align}
where $a \sha b$ denotes the ``shuffle'' of the lists $a$ and $b$, see e.g. \cite{Duhr:2012po, Ablinger:2013cf}.

Many additional relations exist, such as the equivalence to the multiple polylogarithms which are defined as iterated sums, the ``stuffle relation'' between the multiple polylogarithms, various inversion and duplication relations, and many more \cite{Borwein:1999aa, Vollinga:2004sn, Goncharov:2010jf, Duhr:2012po, Ablinger:2013cf, Duhr:2014woa, Panzer:2014caa, Frellesvig:2016ske} which are beyond the scope of these proceedings.

The scope of the current project is, however, slightly different. Where the relations mentioned above relate GPLs to each other, our goal is to express them in terms of some minimal set of simpler functions. Such a minimal set was conjectured by Goncharov as described in ref. \cite{Duhr:2012po}, and specifically is was conjectured that all GPLs of weight\footnote{The weight of a GPL is defined as the number of iterated integrals, so e.g. the GPL given by eq. \eqref{eq:gdef} has weight $n$.} $\leq 4$, which is all that is needed at the level of two-loop Feynman integrals, can be expressed in terms of the functions
\begin{align}
\log(x) \;,\;\; \Li_2(x) \;,\;\; \Li_3(x) \;,\;\; \Li_4(x) \;,\;\; \Litt(x,y).
\end{align}
Here the $\Li_n(x)$ denote the classical polylogarithms, which are defined recursively as
\begin{align}
\Li_n(x) &= \int_{0}^x \! \frac{\id z}{z} \, \Li_{n-1}(z) \;\;\;\;\;\; \tekst{with} \;\;\;\;\;\; \Li_1(x) = -\log(1-x),
\end{align}
and the function $\Litt$ may be expressed in a similar way as
\begin{align}
\Litt(x,y) &= \int_{0}^x \! \frac{\log \big( z/x \big) \Li_2 \big( y z \big)}{z-1} \id z.
\end{align}

It is believed by the authors that most people in the mathematical as well as in the physical communities considered this conjuncture to be true. In any case a lot of expressions for physical quantities such as scattering amplitudes and Wilson loops, expressible in terms of GPLs, have been reexpressed in terms of this minimal set using relatively recently developed methods with names such as ``symbols'' and co-products \cite{Goncharov:2010jf, Duhr:2012po, Duhr:2012fh}. Yet explicit expressions for a general GPL in terms of this minimal basis, valid everywhere in complex phase space, was not known (or at least not written down) and one of the goals of this project was to remedy this, up to weight four.

\section{Reduction}
\label{sec:reduction}

\begin{figure}[ht]
\centering
\includegraphics[width=0.3\textwidth]{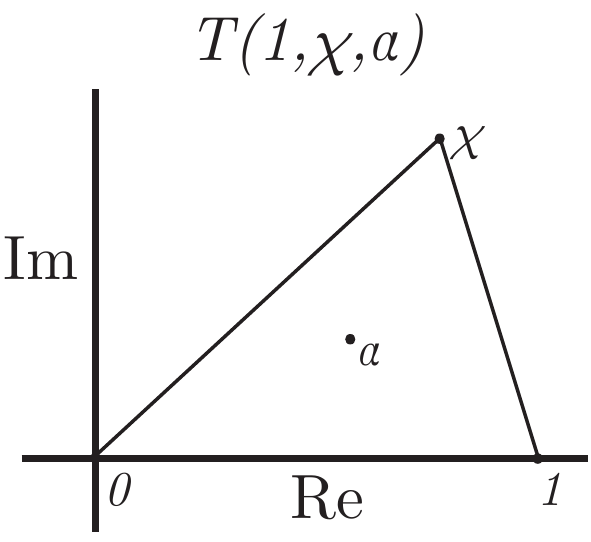}
\caption{This figure shows the triangle referred to by the function $T(1,\chi,\alpha)$. The function evaluates to one when $\alpha$ is inside the triangle, and to zero otherwise.}
\label{fig:triangle}
\end{figure}

At weight one the expression for a general GPL is easily seen from eq. \eqref{eq:gdef} to be
\begin{align}
G(a,x) &= \log(1 - x/a).
\end{align}
We will here show the derivation in full detail for the corresponding expression at weight two, where we assume that $a$, $b$, and $x$ are all different and non-zero. (If that is not the case, the derivation and the result are similar but simpler).
\begin{align}
G(a,b,x) &= G(a/b,1,x/b) \define G(1-\alpha,1,1-\chi) \;\, = \;\, \int_0^{1-\chi} \!\!\!\! \frac{G(1,z)}{z-(1-\alpha)} \id z \\
&= \int_0^{1-\chi} \! \frac{G(0,1-z)}{z-(1-\alpha)} \id z \;\, = \; \int_1^{\chi} \! \frac{G(0,y)}{y-\alpha} \id y \\
&= G(\alpha, 0; \chi) - G(\alpha, 0; 1) + 2 \pi i \, G(0,\alpha) \, \tekst{sgn}(\alpha) T(1,\chi,\alpha) \\
&= G(\alpha, \chi) G(0; \chi) - G(0,\alpha; \chi) - G(\alpha; 1) G(0; 1) \nn \\
& \;\;\; + G(0,\alpha;1) + 2 \pi i \, G(0;\alpha) \, \tekst{sgn}(\alpha) T(1,\chi,\alpha) \\
&= \log(1-\chi/\alpha) \log(\chi) + \Li_2(\chi/\alpha) - \Li_2(1/\alpha) + 2 \pi i \log(\alpha) \tekst{sgn}(\alpha) T(1,\chi,\alpha)
\end{align}
where the last step used the relation $G(0,a,x) = -\Li_2(x/a)$.
The function $T(x,y,z)$ is defined to be one if the complex $z$ is inside the triangle in the complex plane formed by $0$, $x$, and $y$, and zero otherwise, see fig. \ref{fig:triangle}.

At weights three and four the derivation is similar. The first step, i.e. the mapping of e.g. $G(a,b,c,d;x)$ to $G(\alpha,\beta,\gamma,0;\chi)$ can be done recursively as above, see refs. \cite{Vollinga:2004sn, Frellesvig:2016ske}. The remaining step unto the minimal basis gets progressively more difficult for each weight, requiring the systematic use of all shuffle and stuffle relations, in addition to a number of non-trivial integral relations. The result at weight three, for the general $G(a,b,c;x)$, takes up about one page, and the corresponding expression for $G(a,b,c,d;x)$ at weight four, would take up about 20 pages. For the specific expressions and more details of the derivations, see ref. \cite{Frellesvig:2016ske} and its ancillary files.

The mentioned reductions up to weight four, are implemented as a Mathematica replacement rule called {\tt gtolrules}, which is included as an ancillary file to ref. \cite{Frellesvig:2016ske}.

\section{Evaluation}
\label{sec:evaluation}

Having reduced the GPLs to $\Li_n$ and $\Litt$, it becomes desirable to have a quick algorithm for the evaluation of those functions. The implementation of such an algorithm was the second goal of this project. The classical polylogarithm $\Li_n$ is fairly well studied, and fast methods for its evaluation are described in e.g. refs. \cite{Crandall, Vollinga:2004sn}, and in ref. \cite{Frellesvig:2016ske} we described a combined and optimized version of those methods. Yet here we will describe a more primitive evaluation method, as that is the one that generalizes to the case of $\Litt$.

$\Li_n$ may be expressed by the sum
\begin{align}
\Li_n(x) &= \sum_{i=1}^{\infty} \frac{x^{i}}{i^n}
\end{align}
which converges whenever $|x| \leq 1$. In the opposite case one might use the inversion relation for $\Li_n$
\begin{align}
\Li_n(x) = (-1)^{n-1} \Li_n \! \left( \tfrac{1}{x} \right) - \tfrac{1}{n!} \log^n(-x) + 2 \sum_{r=1}^{\floor{\tfrac{n}{2}}} \frac{\log^{n-2r}(-x)}{(n-2r)!} \left( 2^{1-2r} - 1 \right) \zeta(2r)
\label{eq:lininversion}
\end{align}
to map to an $\Li_n$ function with the argument inside the convergent region. An thus we are in principle able to evaluate $\Li_n(x)$ for any argument.

\begin{figure}[ht]
\centering
\includegraphics[width=0.8\textwidth]{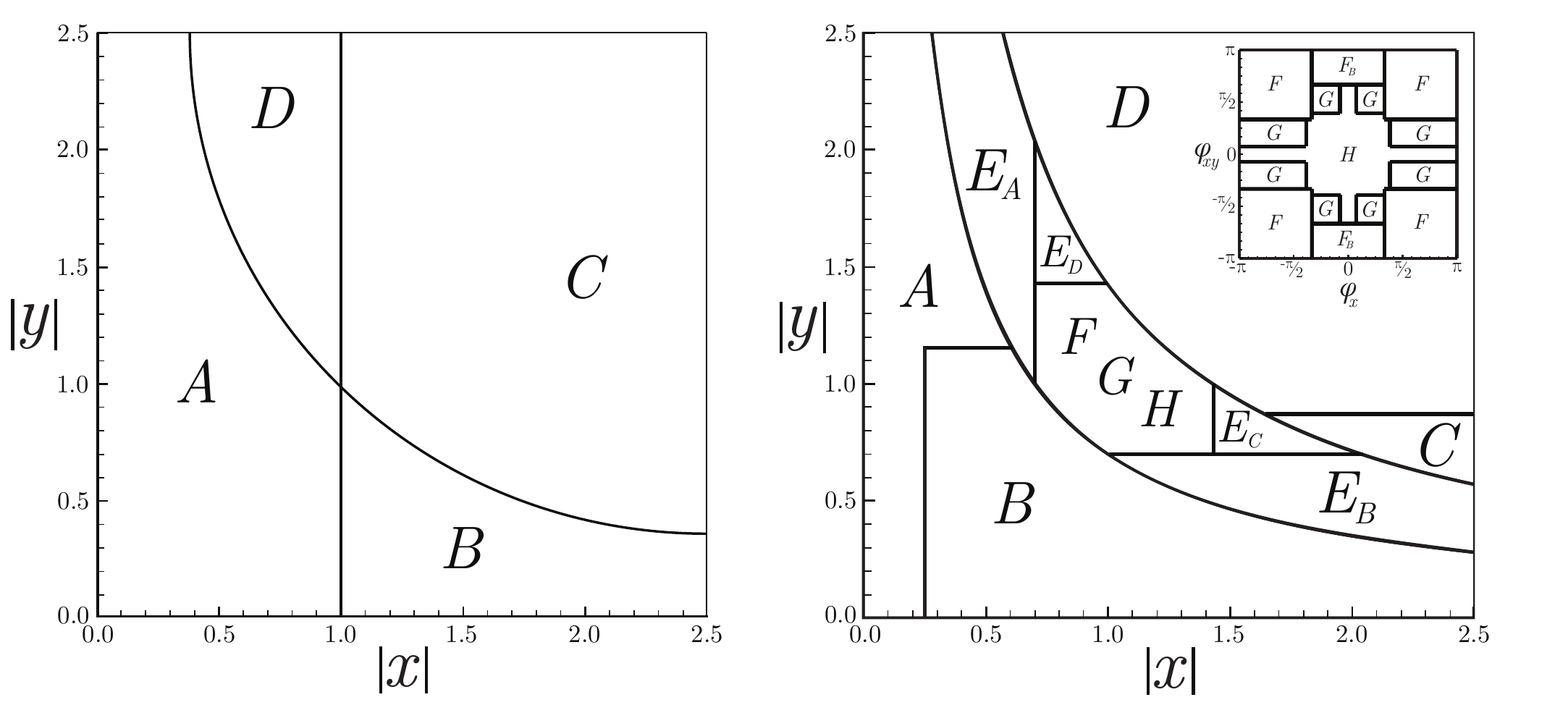}
\caption{This figure describes the regions into which the evaluation of $\Litt(x,y)$ is split. The left figure shows the regions as they are described in the main text, while the right figure shows the regions used by the actual implementation in ref. \protect\cite{Frellesvig:2016ske}. `A' denotes the region of convergence of the sum of eq. \protect\eqref{eq:li22fast}, `B' the region requiring the use of the stuffle-relation \protect\eqref{eq:li22stuffle}, `C' the region of the inversion relation \protect\eqref{eq:li22inversion}, and `D' a region requiring the use of both stuffle and inversion. `E', `F', `G', and `H', refer to regions where different algorithms were used, as described in ref. \protect\cite{Frellesvig:2016ske}.}
\label{fig:li22}
\end{figure}

$\Litt$ is given as a similar sum
\begin{align}
\Litt(x,y) &= \! \sum_{i>j>0}^{\infty} \frac{x^i \, y^j}{i^2 \, j^2}
\label{eq:li22fast}
\end{align}
which converges whenever $|x| \leq 1$ and $|xy| \leq 1$.
For cases where $|xy|>1$ we may use the inversion relation
\begin{align}
\Litt(x,y) &= \Litt \big( \tfrac{1}{x}, \tfrac{1}{y} \big) - \Li_4(xy) + 3 \Big( \Li_4 \big( \tfrac{1}{x} \big) + \Li_4(y) \Big) + 2 \Big( \Li_3 \big( \tfrac{1}{x} \big) - \Li_3(y) \Big) \log(-xy) \nn \\
& \; + \Li_2 \big( \tfrac{1}{x} \big) \left( \frac{\pi^2}{6} + \frac{\log^2(-xy)}{2} \right) + \frac{1}{2} \Li_2(y) \Big( \log^2(-xy) - \log^2(-x) \Big)\,,
\label{eq:li22inversion}
\end{align}
and when $|x|>1$ we may use the stuffle-relation
\begin{align}
\Litt(x,y) &= -\Litt(y, x) - \Li_4(xy) + \Li_2(x) \Li_2(y)\,,
\label{eq:li22stuffle}
\end{align}
and together these relations allow for the mapping of $\Litt(x,y)$ of any set of arguments into the convergent region, see fig. \ref{fig:li22}.

It is easily realized that the sum of eq. \eqref{eq:li22fast} is extremely slowly converging close to $|xy|=1$. For that reason the authors developed a number of alternative algorithms for use in that region. A description of the exact nature of these algorithms is beyond the scope of these proceedings, see ref. \cite{Frellesvig:2016ske} for a more thorough explanation.

A {\tt C++} implementation denoted {\tt lievaluate} of $\Li_n$ and $\Litt$ using the algorithms mentioned above, was included as ancillary files to ref. \cite{Frellesvig:2016ske} alongside code enabling the linking of {\tt lievaluate} from {\tt Mathematica}. The authors note that since the publication of ref. \cite{Frellesvig:2016ske}, an alternative {\tt C++} implementation of these algorithms have been published as ref. \cite{Kirchner:2016hmb}.

\section{Discussion}
\label{sec:discussion}

The reductions described in section \ref{sec:reduction} made no use of the previously mentioned ``symbols'' or coproducts which have been driving a lot of the recent developments in the study and use of GPLs. This was a conscious decision on the side of the authors, as the symbols catch the algebraic but not the analytic parts of the relations they describe. As the goal was to derive relations valid everywhere in the complex phase space, irrespectively of the branch-cuts of the GPL and $\Li$ functions, a symbol-based approach was considered unsuited. Yet there are developments of the use of the coproduct which takes steps towards capturing also the analytic properties of the functions \cite{Abreu:2014cla}, and if one wanted to find the explicit reductions of GPLs at higher weights (such as the 5 and 6 which are needed for three-loop computations), perhaps such an approach would be desirable or even necessary.

\acknowledgments{%
The work of HF is supported by the European Commission through the HiggsTools Initial Training Network PITN-GA-2012-316704.

CW and DT were supported primarily by the Research Funding Program ARISTEIA, HOCTools (co-financed by the European Union (European Social Fund ESF) and Greek national funds through the Operational Program ``Education and Lifelong Learning'' of the National Strategic Reference Framework (NSRF)).
}

\bibliographystyle{ieeetr}
\bibliography{biblio}

\end{document}